\documentclass[pra,floatfix,twocolumn,showpacs,amsmath,amssymb,final]{revtex4}
\usepackage{graphicx} 
\usepackage{dcolumn} 
\usepackage{bm} 
\usepackage[pdftex,colorlinks,bookmarks=false,citecolor=blue,linkcolor=blue,urlcolor=blue]{hyperref}
\usepackage{color}

\begin{document}

\title{Spatial noise correlations of a chain of ultracold fermions - A numerical study}

\author{Andreas L\"uscher}
\affiliation{Institut Romand de Recherche Num\'erique en Physique des Mat\'eriaux (IRRMA), EPFL, CH-1015 Lausanne, Switzerland}
\author{Andreas M. L\"auchli}
\affiliation{Institut Romand de Recherche Num\'erique en Physique des Mat\'eriaux (IRRMA), EPFL, CH-1015 Lausanne, Switzerland}
\author{Reinhard M. Noack}
\affiliation{Fachbereich Physik, Philipps-Universit\"at Marburg, D-35032 Marburg, Germany}

\date{\today}
\begin{abstract}
We present a numerical study of noise correlations, i.e.,
density-density correlations in momentum space, in the extended
fermionic Hubbard model in one dimension. In experiments with
ultracold atoms, these noise correlations can be extracted from
time-of-flight images of the expanding cloud. 
Using the density-matrix renormalization group method to investigate
the Hubbard model at various fillings and interactions, we confirm
that the noise correlations contain full
information on the most important fluctuations
present in the system. We point out the importance of the sum rules
fulfilled by the noise correlations and show that they yield
nonsingular
structures beyond the predictions of bosonization approaches. Noise
correlations can thus serve as a universal 
probe of order and
can be used to characterize the many-body states of cold atoms in optical
lattices. 
\end{abstract}


\maketitle

\section{Introduction\label{sec:introduction}}
Recent advances in methods for trapping and controlling ultracold
atoms have opened up the 
promising possibility of directly simulating strongly interacting
many-body Hamiltonians.
These experimental systems can be used to engineer and analyze models
that lie beyond the scope of present analytical and
numerical methods, thus potentially shedding new light on
fundamental quantum many-body
problems~\cite{Bloch05,Jaksch05,Morsch06,Lewenstein06,Bloch07} such as the
the nature of the mechanism for high-$T_{c}$
superconductivity or whether a true spin liquid can be realized. 
Such systems can be formed by trapping atomic gases
in one-, two-, or three-dimensional optical lattices.
The interactions can be accurately tuned by 
adjusting external fields. In a similar spirit, it has recently been proposed to simulate strongly correlated systems in experiments by studying the dynamics of polaritons in arrays of electromagnetic cavities, see Ref.~\onlinecite{Hartmann06}.

The techniques for manipulating 
ultracold atoms are fairly advanced and have already enabled a
broad range of astonishing systems, such as a superfluid, a Mott
insulator~\cite{Greiner02,Stoferle04}, a strongly interacting Fermi gas~\cite{Ohara02} or also
mixtures of bosonic and fermionic gases~\cite{Ospelkaus06}. However, 
the subsequent analysis of their properties has proven to be difficult. 
In view of the application to solid state problems,
it is crucial to have tools at hand that can accurately describe the
engineered state, preferably by extracting the correlation functions of
the atomic gas. A recent proposal~\cite{Altman04} for a universal
probe of correlations suggested measuring the shot noise in
time-of-flight images of the expanding cloud of atoms after
release from the trap. This method is based on the fact that, after a
long enough time of flight $t$, the density distribution 
of the expanding cloud
becomes proportional to the momentum distribution in the
interacting system~\cite{Roth03,Altman04}, 
\begin{equation*}
\left\langle n\left({\bf r}\right)_{t} \right\rangle \propto 
\frac{m}{\hbar t}\left\langle n_{{\bf q}\left(r\right)} \right\rangle \ ,
\end{equation*}
with momentum ${\bf q}\left({\bf r}\right)=m {\bf r}/\left(\hbar
t\right)$ for an atom of mass $m$.
The noise in the image-by-image statistics is governed by higher order
correlations of the initial state
\begin{align} \label{eq:Grealspace}
G_{\sigma\sigma'}\left({\bf r},{\bf r'}\right)&= \left\langle n_{{\bf q}\left(r\right)}n_{{\bf q}\left(r'\right)} \right\rangle 
- \left\langle n_{{\bf q}\left(r\right)} \right\rangle \left\langle n_{{\bf q}\left(r'\right)} \right\rangle \ ,
\end{align}
where $\sigma$ is an internal quantum number, e.g., the spin, that
allows  different states to be distinguished. 
By analyzing the shot noise in several mean-field states, Altman 
{\it  et al.}~\cite{Altman04} showed that the presence of a particular
order leaves a very distinctive fingerprint on the noise correlations,
e.g., due to superconductivity or spin order. 
On the experimental side, this quantity has already been measured on
several occasions, in both fermionic and bosonic cold atomic gases,
i.e., in bosonic Mott insulators~\cite{Folling05,Spielmann06}, fermionic
superfluids~\cite{Greiner05}, and band insulators~\cite{Rom06}.

In the following, we will
treat one-dimensional (1D) fermionic
systems on an optical lattice and will concentrate on the noise
correlations of the lattice model itself: 
\begin{align} \label{eq:G}
G_{\sigma\sigma'}\left(k,k'\right) &= 
\left\langle n_{k,\sigma} n_{k',\sigma'}\right\rangle -
 \left\langle n_{k,\sigma}\right\rangle \left\langle n_{k',\sigma'}\right\rangle \ , \nonumber \\
G\left(k,k'\right) &= \sum_{\sigma,\sigma'} \ G_{\sigma\sigma'}\left(k,k'\right) \ .
\end{align}
Here the brackets denote the ground state expectation value.
Shortly after the pioneering analysis of Ref.~\onlinecite{Altman04},
which was based on mean-field calculations, Mathey {\it et al.}~\cite{Mathey05}
analyzed the shot noise for a 1D Tomonaga-Luttinger (TL) liquid within
a bosonization approach, allowing them to explore the
momentum-dependence around opposite Fermi points, i.e., around
$k\approx k_{F}$ and $k' \approx -k_{F}$. 
In these TL liquids, different types of order
compete, leading to a rich structure in the noise correlations. 
In the present paper, we employ the density-matrix renormalization group
(DMRG) method~\cite{White92,Schollwoeck05} to study the noise
correlations in the 1D extended Hubbard model. 
This numerical approach
allows us to go beyond the Luttinger theory and to uncover 
the full set
of features that are contained in the noise correlations within the
entire Brillouin zone. In the vicinity of opposite Fermi points, we
find perfect agreement with the analytical predictions of
Ref.~\onlinecite{Mathey05}. 

The remainder of the paper is organized as follows: 
In Sec.~\ref{sec:noiseproperties}, we discuss general properties of
noise correlations, independent of the microscopic model. The extended
Hubbard model considered in this work is then introduced in
Sec.~\ref{sec:hubbardmodel}, together with a summary of the different
(quasi-)orders encountered in our numerical approach. Our main
results, the analysis of the noise correlations for different phases
of the extended Hubbard model, are presented in Sec.~\ref{sec:numerics}. In
Sec.~\ref{sec:experiments}, we make the connection to atomic physics
and discuss experimental issues.
We present our conclusions in
Sec.~\ref{sec:conclusion}.


\section{Properties of Noise Correlations in Fermionic Lattice
    Models\label{sec:noiseproperties}} 
Before considering a specific microscopic model, it is useful to
discuss several general properties of the noise correlation functions
at the lattice level. In the following, we consider the noise
correlations~(\ref{eq:G}) on a periodic lattice with $L$ sites. 
In this case, $k$ denotes the lattice momentum and $\sigma$ describes
an internal quantum number that we associate with the spin, i.e.,
$\sigma \in \{\uparrow,\downarrow\}$. In general, $\sigma$ can denote
a more general flavor or species index;  the statements below remain
valid as long as the density operator
\begin{equation*}
n_{k,\sigma}=c_{k,\sigma}^\dag c_{k,\sigma}^{\phantom{\dag}}
\end{equation*}
can be written as a product of creation and annihilation operators
that satisfy the canonical fermionic commutation rules
\begin{align*}
\{c_{k,\sigma}^{\phantom{\dag}},c_{k',\sigma'}^{\phantom{\dag}}\} &=
\{c_{k,\sigma}^\dag,c_{k',\sigma'}^\dag\}= 0 \ , \\
\{c_{k,\sigma}^{\phantom{\dag}},c_{k',\sigma'}^\dag\} &= \delta_{k,k'}\delta_{\sigma,\sigma'} \ .
\end{align*}
It is convenient to use the Fourier transformation
\begin{equation} \label{eq:fourier}
c_{k,\sigma}^{\phantom{\dag}}=\frac{1}{\sqrt{L}} \sum_{l} e^{i k l}c_{l,\sigma}^{\phantom{\dag}} \ ,
\end{equation}
so that the creation and annihilation operators in coordinate space
also obey the standard commutation relations. Under these assumptions,
$G$ satisfies the following exact statements:
\begin{enumerate}
\item{{\em Bounds:} For all $k,k'$ and $\sigma,\sigma'$, $G$ is uniformly bounded
\begin{equation} \label{eq:Gbounds}
\left|G_{\sigma\sigma'}\left(k,k'\right)\right| \le \frac{1}{4} \ ,
\end{equation}
independent of the system size, i.e., $G_{\sigma\sigma'}(k,k')$ itself
cannot diverge 
with system size.
}
\item{{\em Sum rules:} In a system in which the number of particles of
  every species is conserved, the noise correlations satisfy the sum
  rule 
\begin{equation} \label{eq:sumrule}
\sum_{k,k'} G_{\sigma\sigma'}\left(k,k'\right) = 
\left\langle N_{\sigma} N_{\sigma'} \right \rangle - \left\langle N_{\sigma} \right\rangle 
\left\langle N_{\sigma'} \right \rangle = 0 \ ,
\end{equation}}
where $N_{\sigma} = \sum_k n_{k,\sigma}$. 
\item{{\em Equal-spin momentum diagonal:}  Along the momentum diagonal
  $k=k'$, the equal spin $\sigma=\sigma'$ shot noise is given by  
\begin{equation} \label{eq:Gdiagonal}
G_{\sigma\sigma}(k,k)=\langle n_{k,\sigma} \rangle - \langle n_{k,\sigma} \rangle^2 \ge 0 \ .
\end{equation}
This part of the noise correlation function is therefore completely determined by the momentum distribution function $\langle n_{k,\sigma} \rangle$ itself and does {\em not} contain any other information.}
\end{enumerate}

\begin{table*}
\centering
\begin{minipage}[t]{0.75\linewidth}
\begin{tabular}{c|c|c|c|c|c}
Case & Couplings ($t{=}1$) & $x$ & LRO & QLRO & Strong-coupling limit\\ \hline
\ref{sec:CDW} & $U{=}10,\ V{=}10$  & $1/2$ & CDW &  $-$  & \includegraphics[width=2cm]{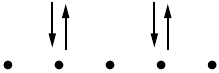} \\[3mm]
\ref{sec:SDW} &$U{=}10$& $1/2$ & $-$ & SDW,pCDW &  \includegraphics[width=2cm]{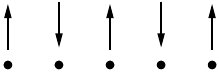} \\[3mm]
\ref{sec:BOW} &$U{=}10,\ t'{=}0.7$  & $1/2$ & pCDW & $-$ &  \includegraphics[width=2cm]{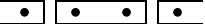} \\[7mm]
\ref{sec:SDWCDW} &$U{=}10,\ V{=}10$  & $1/4$ & CDW $(4k_F)$& SDW & \includegraphics[width=2cm]{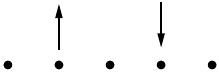} \\[2mm]
\ref{sec:SS} &$U{=}{-}10$ & $3/8$ & $-$& SS [CDW] & \includegraphics[width=2cm]{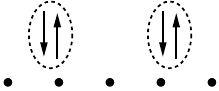} \\[7mm]
\ref{sec:TL} &$U{=}10$ & $3/8$ & $-$&  SDW,CDW [SS,TS] &
\end{tabular}
\end{minipage}
\caption{\emph{Dominant correlations for the parameter sets studied in
    this work,
    supplemented with schematic pictures of the ground states.
    LRO stands for true long-range order, while QLRO
    denotes algebraic correlations (quasi-long-range order). We
    encounter charge-density waves (CDW), spin-density waves (SDW),
    bond-order waves (pCDW, where the kinetic energy is modulated),
    singlet superconductivity (SS) and triplet superconductivity
    (TS). The brackets indicate the subdominant correlation functions,
    i.e., ones which decay with a faster power law than the dominant
    correlations. The momentum of the particle-hole correlations is
    $2k_F=2\pi x$, unless stated otherwise. The pairing correlations
    are for pairs with zero total momentum.
}} \label{tab:cases}
\end{table*}

The basic idea behind the use of density-density correlations as a
universal probe of the system lies in the simple fact that they contain
important parts of various particle-hole {\em and} particle-particle
correlation functions. Quite generally, a particle-hole scattering
operator with net momentum $q$ and form factor $f(k)$ can be written as 
\begin{equation}
\label{eq:ph_formfactor}
\mathcal{O}^{\dag}_\mathrm{p{-}h}(f,q)=\frac{1}{\sqrt{L}}\sum_{k,\alpha,\beta}\ 
f(k)_{\alpha\beta}\ c_{k+q,\alpha}^\dag c_{k,\beta}^{\phantom{\dag}} \ ,
\end{equation}
with 
\begin{equation*} 
f(k)_{\alpha\beta} = 
\begin{cases}
f^\text{CDW}(k) \delta_{\alpha\beta} \\
\frac{1}{2} \mathbf{f}^\text{SDW}(k) \cdot \bm{\sigma}_{\alpha\beta} \ ,
\end{cases}
\end{equation*}
for generalized charge-density waves (CDW) and spin-density waves
(SDW), respectively~\cite{Nayak00}. Here $\bm{\sigma}$ are the Pauli
matrices. This ensemble of CDW and SDW operators encompasses a large
variety of particle-hole instabilities, such as conventional charge
and spin-density waves, bond-order waves, staggered flux phases, as
well as spin nematic states~\cite{Nayak00}. In 1D fermionic systems,
the strongest instabilities in the particle-hole sector often occur at
$q=2 k_F$. However, in some cases, the $q\rightarrow0$ correlations can
also become strong, for example, close to a ferromagnetic transition,
or when longer-range interactions induce important correlations at
wave vectors different from $2k_F$. Analogously, in the
particle-particle sector, a scattering operator of particle pairs with
total momentum $q$ and form factor $f$ reads 
\begin{equation}
\label{eq:pp_formfactor}
\mathcal{O}^{\dag}_\mathrm{p{-}p}(f,q)=\frac{1}{\sqrt{L}}\sum_{k,\alpha,\beta}\ 
f(k)_{\alpha\beta}\ c_{-k+q,\alpha}^\dag c_{k,\beta}^{\dag} \ .
\end{equation}
Here the dependence of the form factor on the spin indices
$\alpha,\beta$ determines whether the operator describes spin-singlet
(SS) or spin-triplet (TS) superconducting pairing. In the case of SS,
$f$ is an even function in the momentum $k$, whereas for TS, $f$ is
odd in $k$. In the following, we restrict our discussion to the most
common case of pairs with total momentum $q=0$, but we
remark that
the Fulde-Ferrell-Larkin-Ovchinnikov (FFLO)~\cite{Casalbuoni04}  
state appearing in the current discussion on the superfluid pairing
properties in population-imbalanced Fermi gases would require a finite
$q$. 

The strongest correlations can be identified by comparing the structure
factors of possible competing orderings. With the above definitions of
the scattering operators, the associated structure factors can be
expressed as
\begin{equation}\label{eq:S}
S_{\xi}(f,q) = \left\langle \mathcal{O}_{\xi}^{\dag}(f,q) 
\mathcal{O}_{\xi}(f,q) \right\rangle \ .
\end{equation}
In the case of true long-range order, the structure factor taken at
the ordering wave vector $k$ would grow linearly with the system size and
would thus diverge in the thermodynamic limit. For the quasi-long-range
order encountered in many 1D systems, in which the asymptotic decay
of the correlation functions has the form of a 
power law (algebraic decay), the corresponding structure factors can
show similar power-law 
divergences, with exponents that depend on the value of the interactions in the
microscopic model. 
To see the tight
correspondence between the structure
factor~(\ref{eq:S}) and the noise correlation function~(\ref{eq:G}),
it is 
illuminating to rewrite the operators in a slightly different
way. For instance, the four-body operators of the particle-hole
structure factor 
that contribute to the noise correlations read
\begin{multline*}
\left. f(k)^*_{\alpha\beta} f(k')_{\alpha' \beta'} c_{k+q,\alpha}^\dag
c_{k,\beta}^{\phantom{\dag}} c_{k',\beta'}^{\dag}
c_{k'+q,\alpha'}^{\phantom{\dag}}    
\right|_{\substack{k'=k \\ \alpha=\alpha' \\ \beta=\beta'}} \\ = 
- \left| f(k)_{\alpha\beta}\right|^2 c_{k,\beta}^\dag
c_{k,\beta}^{\phantom\dag} 
c_{k+q,\alpha}^\dag c_{k+q,\alpha}^{\phantom\dag} + \ldots \ ,
\end{multline*}
where the expression on the second line is easily recognized as one term
of the noise correlations. The additional terms are
two-body operators
that arise due to commutation rules. From this ``identification'', we
draw three important conclusions: First, 
only the modulus of the form factors $f(k)$ enters the noise correlations. Two
types of order that differ only by a phase factor are thus difficult
to distinguish by inspecting only the vicinity of the Fermi
points. For instance, a bond-order-wave state, i.e., a bond-centered
CDW with $p$-wave character, and a conventional site-centered $s$-wave
CDW lead to similar features in the shot noise. Second, since the
noise correlations are bounded by Eq.~(\ref{eq:Gbounds}), they cannot
diverge.
However, we expect any
divergences present in the structure factors, properly rescaled, to 
also appear in the
rescaled noise correlations $LG$, where $L$ is the size of the system;
see also Ref.~\onlinecite{Mathey05}. Third, it is interesting to note
that only the nature of a fluctuation, i.e., whether it is particle-hole or
particle-particle-like, determines the sign of the
divergence. According to the analysis above,
particle-hole
correlations enter with a negative sign as dips along 
$k'=k \pm q$, while particle-particle fluctuations give a positive 
contribution and are observed along the anti-diagonal $k'=-k$ of the shot noise. 
This is what is expected intuitively because a particle-hole scattering
process is only effective if the $k$-orbital is occupied in a
different way than the $k\pm q$-orbital, i.e., if they are
anticorrelated. On the other hand, for particle-particle scattering,
the $k$ and the $-k$-orbitals must be both filled or  both empty in
order to be effective, i.e., they must be (positively) correlated.


\section{The extendend Hubbard model\label{sec:hubbardmodel}}

\begin{figure*}
\includegraphics[width=0.8\textwidth,clip]{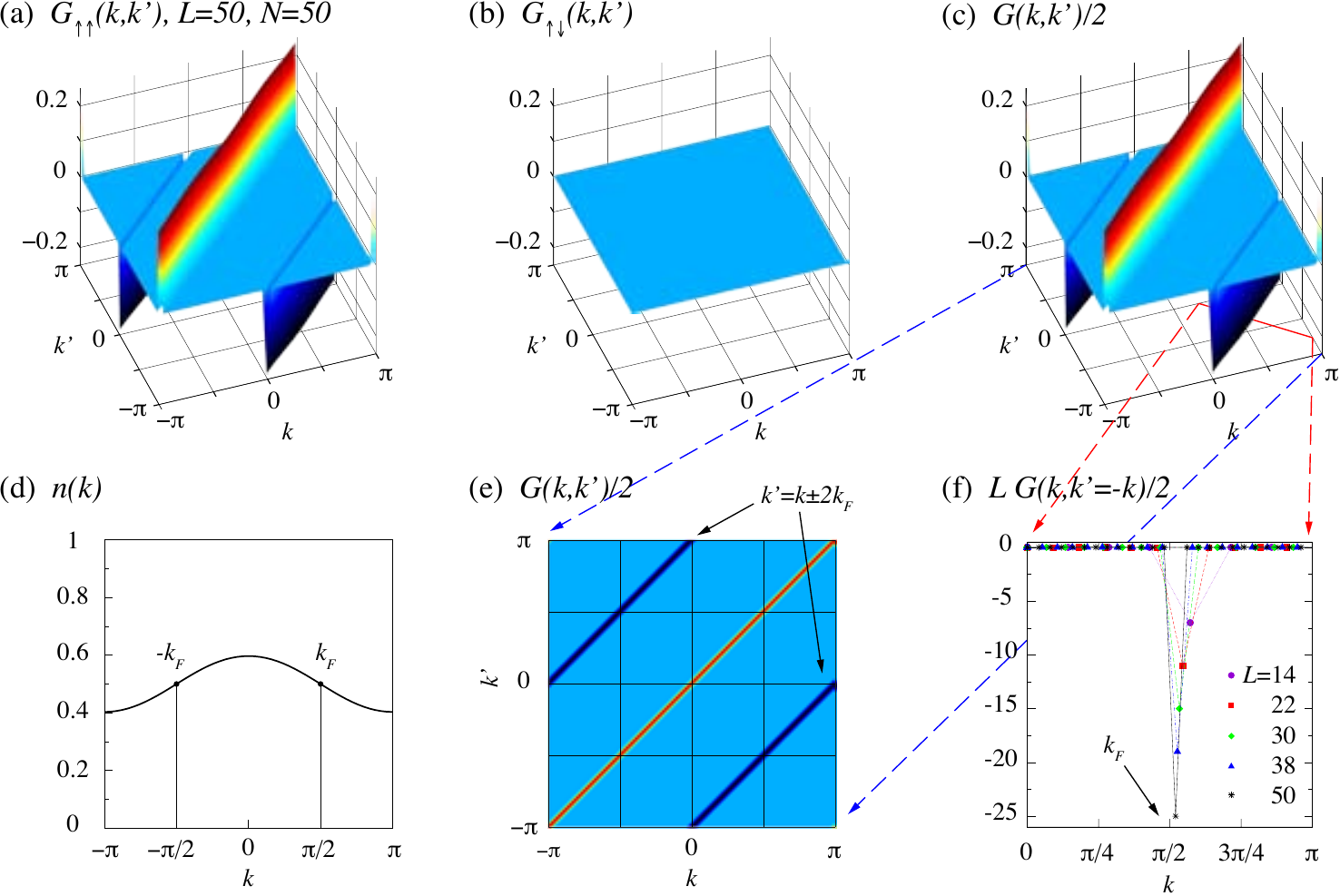}
\caption{\emph{(Color online). Noise correlations (a, b, c, e, f) and momentum
    distribution (d) in the extended Hubbard chain with $U/t=V/t=10$
    at filling $x=1/2$, i.e., $k_{F}=\pi/2$, obtained using
    coordinate-space DMRG calculations and subsequent Fourier
    transformation. The system exhibits an ordered $2k_{F}$-CDW, manifesting itself
    in the pronounced dip along $k'=k\pm2 k_{F}$ in
    $G_{\uparrow\uparrow}$ (a) and vanishing correlations 
    in the $\uparrow\downarrow$-channel (b).
    This can be best seen in the 
    intensity plot of the 
    total noise correlations (e). The rescaled shot noise (f) at 
    $k=k_{F}$, $k'=-k_{F}$ scales linearly with the size of the system
    $L$, indicating true long-range order. Note that because of the closed shell
    configurations used in these calculations, $k_{F}$ is not part of the discrete Brillouin zone.
    The finite-size scaling (f) thus shows paths along $k'=-k+2\pi/L$.
}\label{fig:cdw}}
\end{figure*}
\begin{figure*}
\includegraphics[width=0.8\textwidth,clip]{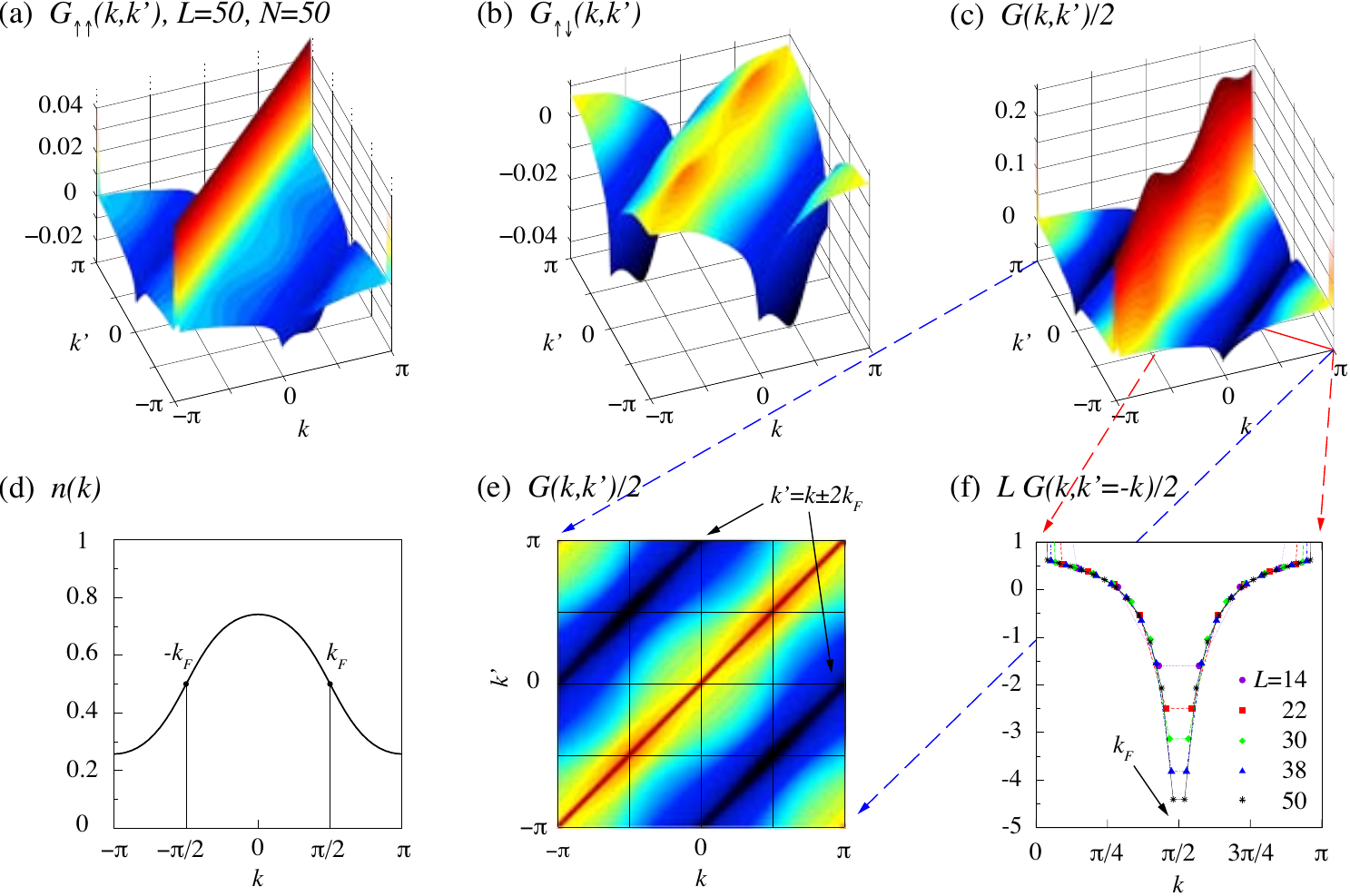}
\caption{\emph{(Color online). Noise correlations (a, b, c, e, f) and momentum
    distribution (d) for the Hubbard chain with $U/t=10$ and $V=0$ at
    half-filling, i.e., $k_{F}=\pi/2$. The system is in the Heisenberg
    regime where only spin-spin correlations are critical because the
    charge sector is gapped. Note that in (a), we have chosen a very
    small upper cutoff to reveal the structure around opposite Fermi
    points. The behavior along the diagonal can be deduced from
    (c). The system exhibits a $2k_{F}$-SDW, which manifests itself in
    both the $\uparrow\uparrow$ (a) and the $\uparrow\downarrow$-channel (b)
    as dips along $k'=k\pm2 k_{F}$. This is well visible in the intensity plot (e).
    In contrast to the case of Fig.~\ref{fig:cdw}, this system only exhibits quasi long-range order. 
    The rescaled shot noise (f) thus grows much more slowly than in the case of true
    long-range order.
}\label{fig:sdw}} 
\end{figure*}

We now consider the extended Hubbard model on a chain with $L$ sites 
and periodic boundary conditions (PBC).
The Hamiltonian of this system reads 
\begin{multline} \label{eq:H}
H=-t \sum_{l,\sigma} \left( c_{l,\sigma}^\dag c^{\phantom{\dag}}_{l+1,\sigma} + \text{H.c.}\right)
+ U \sum_{l} n_{l,\uparrow} n_{l,\downarrow}  \\ 
 -t' \sum_{l,\sigma} \left( c_{l,\sigma}^\dag
 c^{\phantom{\dag}}_{l+2,\sigma} + \text{H.c.}\right) 
+ V \sum_{l} n_{l} n_{l+1} \ ,
\end{multline}
where $c_{l,\sigma}^\dag$ ($c^{\phantom{\dag}}_{l,\sigma}$) creates
(destroys) an electron with spin $\sigma$ on site $l$,  $n_{l,\sigma}
= c_{l,\sigma}^\dag c^{\phantom{\dag}}_{l,\sigma}$ is the occupation
number operator and the parameters $t'$, $U$ and $V$ characterize the
next-nearest-neighbor hopping, the on-site and the nearest-neighbor
interactions, respectively. We focus on the ground state sector with
$N_{\uparrow} = N_{\downarrow}=N/2$, with $N$ the total number of
particles. The filling $x$ is defined as $x=N/(2 L)$ and thus, as
usual, half-filling ($x=1/2$) corresponds to one electron per site on
average. For convenience, we set the hopping amplitude $t=1$ and
express the other interactions in units of $t$. The extended Hubbard
model is $SU(2)$--invariant for all values of the couplings, and the
ground state is a spin singlet for the parameter values considered in
this work. Longitudinal and transverse spin correlations, which appear in
$G_{\uparrow\uparrow}$ and $G_{\uparrow\downarrow}$, respectively, are
therefore identical by symmetry. Furthermore, the continuous $U(1)$
gauge symmetry and the $SU(2)$ spin invariance forbid the appearance
of true superconducting or spin-ordered states in this 1D model.
However, the discrete lattice symmetries or the time-reversal symmetry
can, in principle, be broken, giving rise to commensurate
charge-density or bond-order waves. 
\begin{figure*}
\includegraphics[width=0.8\textwidth,clip]{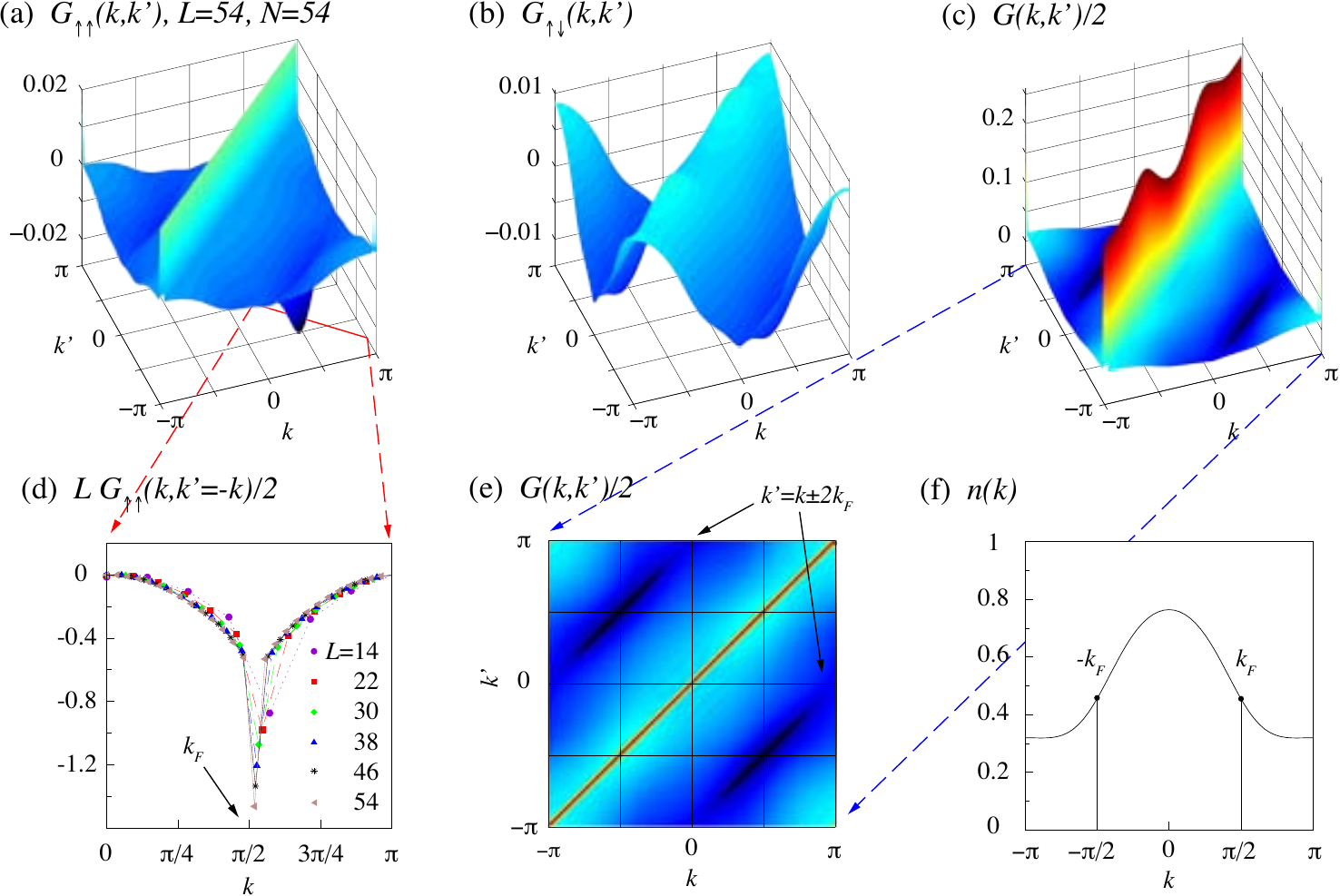}
\caption{\emph{(Color online). Noise correlations (a, b, c, d, e) and momentum
    distribution (f) in the Hubbard chain with $U/t=10$, $t'/t=0.7$,
    and $V=0$ at filling $x=1/2$. The system is in a phase with
    long-range modulations of the kinetic energy.
    In $G_{\uparrow\uparrow}$, (a), a rather sharp dip
    along $k'=k\pm\pi$ can be seen. The finite-size scaling (d)
    illustrates the long-range character of the correlations. Note that 
    because of the closed shell configurations used in these calculations, 
    $k_{F}$ is not part of the discrete Brillouin zone.
    The finite-size scaling (d) thus shows paths along $k'=-k+2\pi/L$.
    In the $\uparrow\downarrow$-channel (b), only a shallow valley can be
    identified along the diagonal $k'=k\pm\pi$. 
    Finite-size scaling (not shown) indicates that these spin
    correlations are short-range. Due to the phase insensitivity, the
    noise correlations alone cannot distinguish between this
    bond-centered and a conventional site-centered CDW, when focussing
    only on a close neighborhood of the Fermi points.}
\label{fig:bow}}
\end{figure*}

For the generic case of repulsive interactions and incommensurate fillings, 
the low-energy sector of the extended Hubbard model is accurately described as a TL
liquid~\cite{Vojt94} with four independent parameters that completely
specify the model: the velocities of the charge and spin excitations,
$v_{\rho}$ and $v_{\sigma}$, respectively, and the correlation
exponents $K_\rho$ and $K_\sigma$. Since the $SU(2)$ symmetry of the
extended Hubbard model implies $K_{\sigma}=1$, the asymptotic behavior
of the correlation functions depends only on $K_{\rho}$, which in
terms depends on the interactions and the filling. In the absence of
interactions, these velocities reduce to the Fermi velocity $v_F$, and
the correlation exponents are equal to $1$. 

The Hamiltonian~(\ref{eq:H}) has been studied extensively in the past
(see, e.g., Ref.~\onlinecite{GiamarchiBook} for an overview). Based on
this knowledge, we have chosen a few interesting parameter sets in the
global phase
diagram~\cite{Schulz90,Bogolyubov90,Frahm90,Mila93,Penc94,Daul98,Daul00}
to investigate the noise correlations, see Tab.~\ref{tab:cases}. The
presentation in Sec.~\ref{sec:numerics} is 
ordered according to increasing complexity of the noise spectrum.


\section{Numerical results\label{sec:numerics}}
In this section, we present our numerical results for the noise
correlations $G$ (\ref{eq:G}) in the extended Hubbard
chain~(\ref{eq:H}) obtained by DMRG calculations in coordinate space
and exact diagonalizations (ED) in momentum
space~\cite{Laeuchli04}. Due to the vast number $\left[{\cal O}(L^4)\right]$ of four-point 
correlation functions that have to
be evaluated in a coordinate space approach, we only consider chains
with up to 56 sites. 
These sizes are, however, sufficient for the purpose of the
present work because the main features of the noise correlations are
already well-established in these systems  
and because the finite-size scaling does not show any ambiguities. 
We have checked the
implementation and the accuracy of the coordinate-space DMRG calculations
against ED results on smaller samples (up to $L=20$ sites) and have found
very good agreement. In addition, our calculations all comply very
well with the sum rule~(\ref{eq:sumrule}). 

The number of fermions $N$ and the system size $L$ are always chosen
so that {\em closed shell}, 
configurations occur, i.e., so that all orbitals below
$k_{F}$ are occupied,  while orbitals above $k_{F}$ are empty. This
condition translates into a fermion number $N=2+4 m$, where $m$ is a
nonnegative integer and hence $L=N/\left(2x\right)$. On the one hand, this
choice eliminates spurious effects arising from open-shell conditions,
but, on the other hand, the Fermi vector $k_{F}$ cannot be part of the
discrete Brillouin zone; only even multiples of it can be
resolved. Nevertheless, the effects we will study will be well-visible in the
vicinity of the Fermi momentum; its behavior for $k \rightarrow \pm
k_{F}$ can easily be deduced. 

\subsection{Ordered charge-density wave\label{sec:CDW}}
Let us start with a simple case that exhibits true long-range order by
choosing a half-filled system ($x=1/2$) with $U/t=10$ and $V/t=10$. In
this case, the charge and the spin sectors are gapped. In the limit
$t=0$, $V/2>U$ is a sufficient condition for the existence of a
charge-density-wave state in which every second site is doubly
occupied, see Tab.~\ref{tab:cases}.
In this strong-coupling limit,
the noise correlations can be calculated analytically using a
site-factorized wave function 
\begin{align} \label{eq:GCDW}
G_{\uparrow\uparrow}\left(k,k' \ne k\right) &=
-\frac{1}{L^2} \sum_{l=1}^{L/\mu} \sum_{l'=1}^L 
e^{i\left(k-k'\right)\left(l-l'\right)} \delta_{l',l+m \mu} \nonumber \\
&\approx -\frac{1}{\mu^2} \delta_{k',k\pm\frac{2\pi}{\mu}} \ , \nonumber  \\
G_{\uparrow\downarrow}\left(k,k'\right) &=0 \ .
\end{align}
Here $\mu$ is the period of the modulation and $m$ an integer. 
We thus expect a signature along $k'=k\pm2\pi/\mu$.  

These arguments are in perfect quantitative agreement with the
numerical results for the shot noise shown in Fig.~\ref{fig:cdw},
which clearly reveal a $2 k_{F}=\pi$ CDW signature in the
$\uparrow\uparrow$-channel and vanishing correlations in the
$\uparrow\downarrow$-channel. Although the featureless momentum
distribution shown in Fig.~\ref{fig:cdw}(e) is an indicator of a
gapped state, it does not reveal its nature. Since we have chosen the
number of particles and the length of the chain to exclude the
Fermi points, a slice along the anti-diagonal $k'=-k$ does not contain
the minimum of the dip. However, along a slightly shifted path along
$k'=-k+2 \pi/L$, as depicted in Fig.~\ref{fig:cdw}(f), it is easy to see
that the minimum of the {\it rescaled} shot noise increases linearly
with the size of the system, in agreement with the strong-coupling
derivation above. In the thermodynamic limit, the rescaled shot noise
thus contains a diverging dip along $k'=k\pm2 k_{F}$. 

\subsection{Algebraically decaying spin-density wave\label{sec:SDW}}

Without the nearest-neighbor interaction $V$, one recovers the
standard Hubbard chain. At half-filling, the charge sector is gapped,
but the spin excitations are gapless. In this case, the correlation
exponent $K_{\rho}$ vanishes and density-density correlations decay
exponentially. Because of the $SU(2)$ invariance, $K_{\sigma}=1$ and
the spin-spin correlations are critical, asymptotically decaying as
$1/r$.  
The low-energy sector of a half-filled Hubbard model is equivalent to
a $S=1/2$ Heisenberg chain. A snapshot of a spin configuration is depicted in
Tab.~\ref{tab:cases}.
The simple analytical strong-coupling analysis
above is equally applicable to spin modulations; the only difference
is that spin correlations enter not only 
$G_{\uparrow\uparrow}$, but also $G_{\uparrow\downarrow}$. The
periodicity of the modulation is again two lattice spacings, leading
to a signal along $k'=k\pm 2 k_{F}$. In contrast to the previous case,
this system does not exhibit true long-range order because the SDW
state would break a continuous symmetry. The ordering tendencies are
accordingly less distinct than in the CDW example with true long-range
order. This is illustrated in the finite-size scaling analysis of the
rescaled shot noise $LG$ in Fig.~\ref{fig:sdw}(f). 

\begin{figure*}
\includegraphics[width=0.8\textwidth,clip]{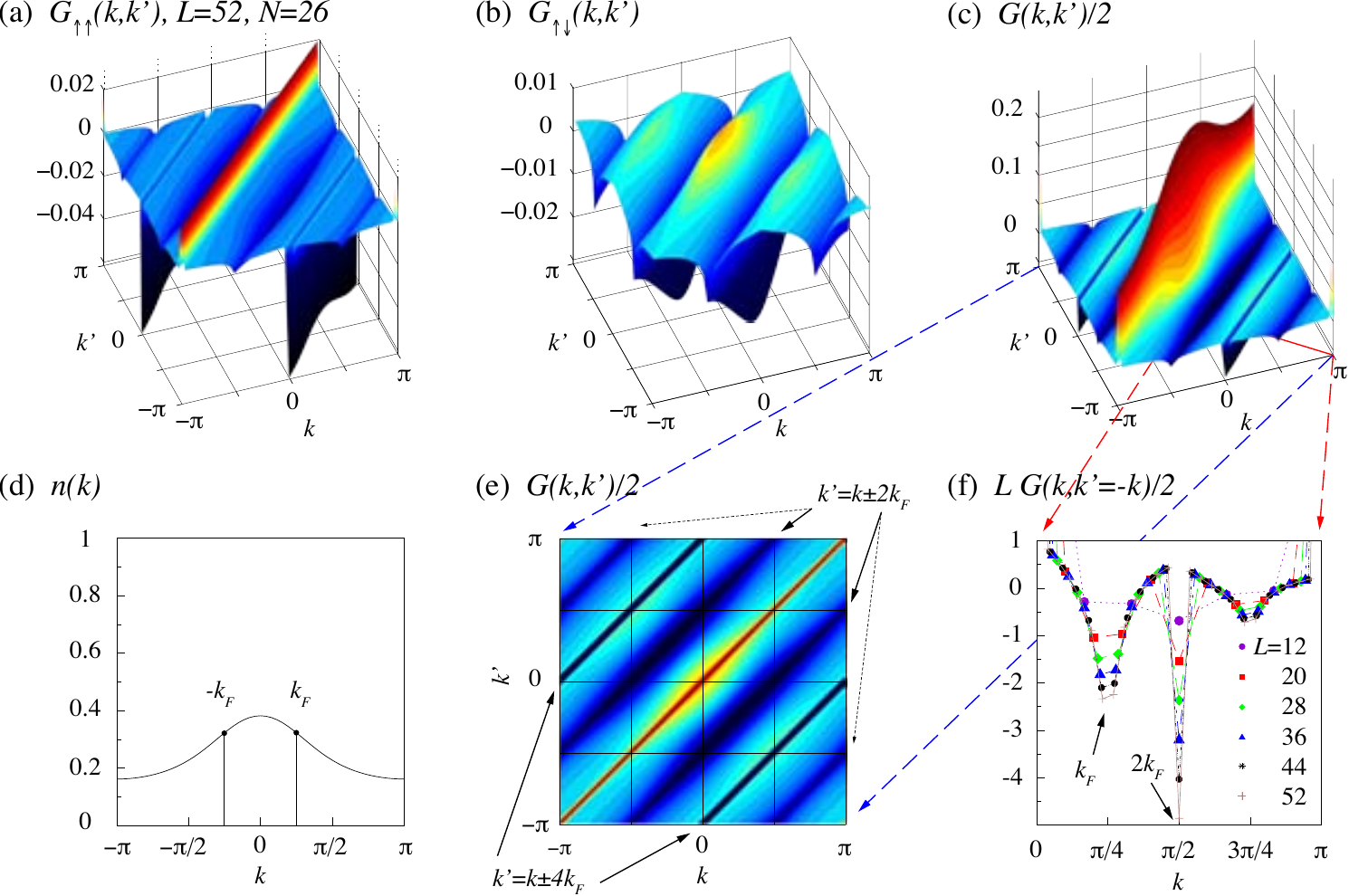}
\caption{\emph{(Color online). Noise correlations (a, b, c, e, f) and momentum
    distribution (d) in the extended Hubbard model with $U/t=V/t=10$
    at quarter-filling, i.e., $k_{F}=\pi/4$. This system is characterized 
    by long-range charge modulations at
    wave-vector $4 k_{F}$ in $G_{\uparrow\uparrow}$ (a) and
    algebraically decaying spin-spin fluctuations at $2 k_{F}$ in
    $G_{\uparrow\uparrow}$ and $G_{\uparrow\downarrow}$ (b).
    The small dips
    along $k'=k\pm3\pi/2$ are simply due to the periodicity of the
    Brillouin zone. A finite-size scaling analysis of the total shot
    noise along the anti-diagonal (f) nicely illustrates the
    difference between true and quasi-long-range
    order.}\label{fig:sdw-cdw}}
\end{figure*}

In addition to the well-known staggered spin correlations, the
Heisenberg chain also has staggered dimer-dimer correlations decaying
with the same power law (disregarding logarithmic corrections). In
fermionic language, these dimer correlations share the same properties
as the bond-order-wave correlations, i.e., are modulations of the kinetic
energy. In terms of higher angular momentum density waves, this is a
$p$-wave CDW. We therefore see a superposition of both the spin and
the kinetic energy correlations in $G_{\uparrow\uparrow}$ along $k'=k\pm2k_F$.  

\subsection{Ordered bond-order wave (dimerized) state\label{sec:BOW}}
To extend the discussion on critical bond-order wave
correlations in the half-filled Hubbard model, it is interesting to
separate the two contributions by driving the system into a dimerized
phase, where the kinetic energy is modulated and the spin-spin
correlations decay exponentially. Such a phase has been reported in
Ref.~\onlinecite{Daul00}, in a study of the Hubbard model ($V=0$) with
an additional next-nearest neighbor hopping $t'$, where a spin gap
opens and the spin correlations become very short-range. In the large
$U$ limit, this model maps onto a frustrated Heisenberg chain, which
is known to be dimerized at the Majumdar-Ghosh point 
$J_2/J_1 = 1/2$~\cite{Majumdar69}. 
In  Ref.~\onlinecite{Daul00}, it has been
shown that this phase persists for finite values of $U$. We have
therefore chosen to investigate the noise correlations for a
half-filled chain with $U/t=10$ and $t'/t=0.7$. 
In a state where the kinetic energy is modulated with period $\mu$, we
expect pronounced correlations of the form  
\begin{equation*}
\left\langle \left(c_{i,\sigma}^{\dag} c_{i+1,\sigma}^{\phantom{\dag}} 
+\text{H.c.}\right)\left( c_{i+\mu,\sigma'}^{\dag} 
c_{i+\mu+1,\sigma'}^{\phantom{\dag}}+\text{H.c.}\right) \right\rangle \ .
\end{equation*}
Tab.~\ref{tab:cases}
depicts such a dimerized state (with modulation
period $\mu=2$ lattice spacings) schematically. 
Writing the noise
correlations~(\ref{eq:G}) in coordinate space, one can convince
oneself (after a bit of algebra) that a bond-order wave leaves a
non-trivial fingerprint only in the $\uparrow \uparrow$-channel along
$k'=k\pm\pi$. 
Due to the particle-hole nature of the fluctuation, the
signature is negative. 
The $\uparrow \downarrow$-channel contains
non-singular contributions of the form $e^{\pm i k} e^{\pm i k'}$. 
This crude approximation is confirmed by our numerical results
presented in Fig.~\ref{fig:bow}. 
The well-pronounced dip along
$k'=k\pm\pi$ in $G_{\uparrow\uparrow}$ is clearly distinguishable from
the shallow surface in $G_{\uparrow\downarrow}$. The finite-size scaling
[Fig.~\ref{fig:bow}(d)] illustrates the long-range character of the
correlations. 
In contrast, the finite-size scaling of 
$G_{\uparrow \downarrow}$ along the same path (not shown) quickly saturates,
showing that the spin correlations are indeed short-range. 
Note that the  noise correlations alone cannot
distinguish between this bond-centered and a conventional
site-centered CDW in 1D due to the phase insensitivity.
\begin{figure*}
\includegraphics[width=0.8\textwidth,clip]{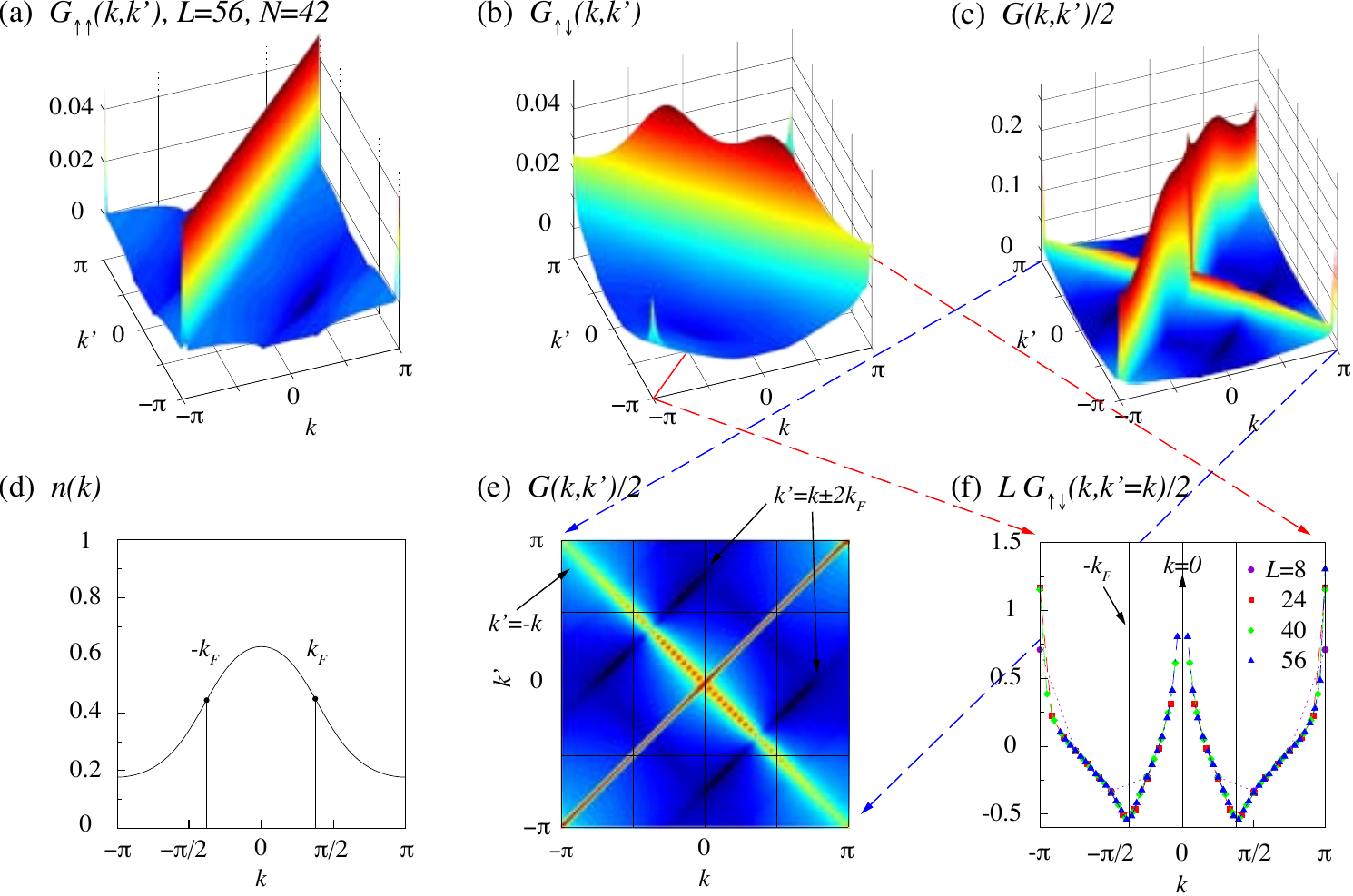}
\caption{\emph{(Color online). Noise correlations (a, b, c, e, f) and momentum
    distribution (d) in the attractive Hubbard chain with $U/t=-10$
    ($V=0$) at filling $x=3/8$. In real space, electrons tend to form
    coherent on-site singlets. In momentum space, this pairing is reflected in dominant particle-particle
    correlations along $k=-k'$ in $G_{\uparrow\downarrow}$ (b). In the
    total shot noise (f), it becomes apparent that the $2k_{F}$-CDW
    fluctuations present in $G_{\uparrow\uparrow}$ (a) are decaying faster
    than the superconducting fluctuations. In contrast to the noise
    correlations, the momentum distribution (d) does not reveal any
    information on the ordering tendencies. Note in (f) the presence of a
    kink-like negative feature at $\pm k_F$, which compensates the positive
    response of the pairing correlations, by virtue of the sum rule (\ref{eq:sumrule}).}
    \label{fig:sc-cdw}} 
\end{figure*}

\subsection{Ordered charge-density wave coexisting with an
  algebraically decaying spin-density wave\label{sec:SDWCDW}}
Another interesting example is a system with both SDW quasi-order
and true long-range CDW ordering at different wave vectors, realized
for a quarter-filled chain with $U/t=V/t=10$. For these parameters,
the model is an insulator with a charge gap, i.e., the correlation
exponent $K_{\rho}=0$. As will be given explicitly in
Eqs.~(\ref{eq:C2}), the spin-spin 
correlations decay as $1/r$, as in the case of the Heisenberg
chain. The strong-coupling ground state is shown in
Tab.~\ref{tab:cases}.
Based on this picture, we expect $2 k_{F}$
spin modulations and $4 k_{F}$ charge modulations to be present. 
This prediction is in excellent agreement with the numerical results
of Fig.~\ref{fig:sdw-cdw}. 
A pronounced CDW response is visible along
$k'=k\pm\pi$ in $G_{\uparrow\uparrow}$ [Fig.~\ref{fig:sdw-cdw}(a)],
while less pronounced SDW ordering can be identified in
$G_{\uparrow\uparrow}$ and $G_{\uparrow\downarrow}$
[Fig.~\ref{fig:sdw-cdw}(b)] along $k'=k\pm\pi/2$. Note that the small
dips along $k'=k\pm3\pi/2$ in Fig.~\ref{fig:sdw-cdw}(b) are just due
to the periodicity of the Brillouin zone. The finite-size scaling
along the anti-diagonal clearly reveals that
charge ordering is much stronger than the tendency to form a SDW, as
can be seen in Fig.~\ref{fig:sdw-cdw}(f).

\subsection{Algebraically decaying singlet superconductivity and 
charge-density wave correlations\label{sec:SS}}

For an attractive on-site interaction ($U/t<0$), the fermions tend to
pair in singlets, leading to a gap in the spin sector, i.e.,  it
requires a finite energy to break up an 
on-site pair because, even in
the thermodynamic limit, spin correlations and triplet superconducting
correlations decay exponentially,  
while singlet superconducting and charge fluctuations exhibit an
asymptotic behavior of the form~\cite{Bogolyubov90}
\begin{align*}
C_{SS}\left(l\right) &= \langle c_{l,\uparrow}^\dag 
c_{l,\downarrow}^\dag c_{0,\downarrow}^{\phantom{\dag}} 
c_{0,\uparrow}^{\phantom{\dag}}\rangle 
\rightarrow \frac{A_{SS}}{\left| l \right|^{1/K_{\rho}}} \ , \\
C_{CDW}\left(l\right) &= \left\langle n_{l} n_{0} 
\right\rangle \rightarrow n^2 
+A_{CDW}\frac{\cos(2k_{F} l)}{\left| l \right|^{K_{\rho}}} \ ,
\end{align*}
with amplitudes $A_{\xi}$ that depend on the values of the
interactions and on the
filling. 
The presence of a spin gap is taken into account by formally
setting the correlation exponent $K_{\sigma}\rightarrow0$. Since
$K_{\rho}>1$ for attractive interactions and away from half filling,
the SS fluctuations are always dominant and the tendency to form a
CDW is subdominant.  Fig.~\ref{fig:sc-cdw} shows the noise
correlations and the momentum distribution [Fig.~\ref{fig:sc-cdw}(d)]
calculated for a Hubbard model with attractive interaction $U/t=-10$
($V=0$) at a filling of $x=3/8$.
Based on Ref.~\onlinecite{GiamarchiBook}, we estimate $K_{\rho}\approx
1.3$. Note that in Fig.~\ref{fig:sc-cdw}(a), we have chosen a very
small cutoff in order to display the much smaller CDW signature along
$k'=k+2 k_{F}$ in $G_{\uparrow\uparrow}$. As expected, SS correlations
are visible along the anti-diagonal in $G_{\uparrow\downarrow}$ and
clearly dominate over the CDW signature; 
this is evident in
the total shot noise, Fig.~\ref{fig:sc-cdw}(c). In the region
accessible to bosonization, i.e., around opposing Fermi points, our
results for the noise correlations agree well with the behavior
illustrated in the last column of Fig.~3 of
Ref.~\onlinecite{Mathey05}.  Note, however, that the sum rule
(\ref{eq:sumrule}) leads to nonsingular structures in
$G_{\uparrow\downarrow}$.  Such structures can be clearly seen in
Fig.~\ref{fig:sc-cdw}(f) as pronounced dips along $k=\pm k_F$. These
dips and valleys are necessary to compensate the strong positive
contributions due to the singlet superconducting correlations and are
not predicted by bosonization.

\begin{figure*}
\includegraphics[width=0.8\textwidth,clip]{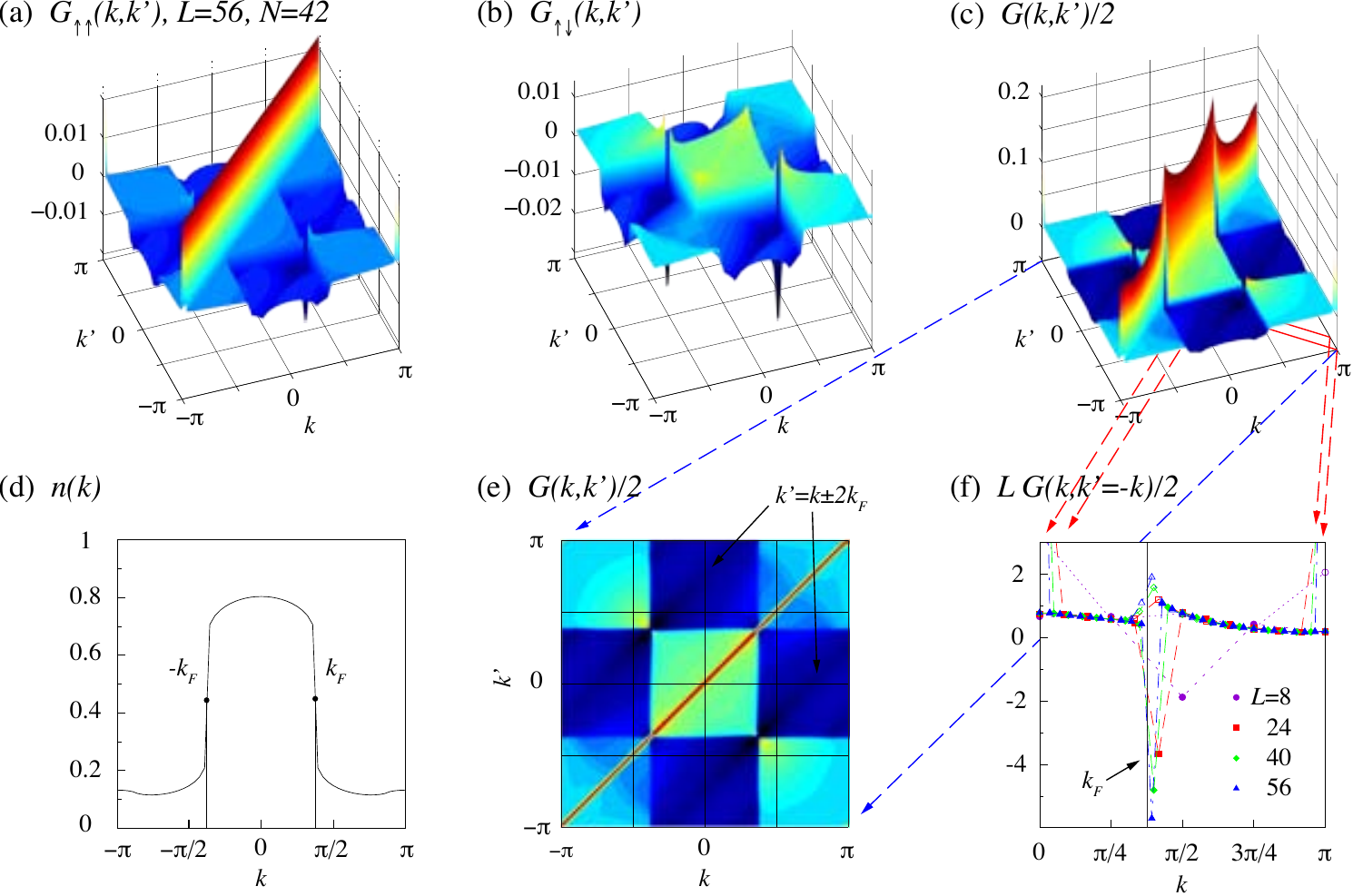}
\caption{\emph{(Color online). Noise correlations (a, b, c, e, f) and momentum
    distribution (d) in the Hubbard model in a metallic phase with
    $U/t=10$ ($V=0$) at filling $x=3/8$. The metallic character is
    reflected by the infinite slope (not jump) singularity of the
    momentum distribution (d) at $k=\pm k_{F}$. This singularity also
    translates itself to the noise correlations. CDW and SDW
    signatures (with a negative sign, due to their particle-hole
    nature) are recognizable along $k'=k\pm2 k_{F}$, while the small
    positive peaks around the Fermi points are due to the pairing
    fluctuations. The chosen parameters correspond to a Luttinger
    liquid correlation exponent
    $K_\rho\approx0.58$}.\label{fig:metal}}
\end{figure*}

\subsection{Metallic Tomonaga-Luttinger liquid\label{sec:TL}}
We take the conventional repulsive Hubbard model at filling $x=3/8$
with $U/t=10$ as an example of a spin-$1/2$ Luttinger liquid with
both gapless charge and spin excitations. From
Refs.~\onlinecite{Schulz90,Ejima05}, we estimate
$K_{\rho}\approx0.58$, while $K_\sigma=1$ due to the $SU(2)$
symmetry. According to Refs.~\onlinecite{Giamarchi89,Frahm90}, the
density-density and spin-spin correlations decay with the same
exponent $1+K_{\rho}$ (neglecting logarithmic corrections)
\begin{subequations} \label{eq:C2}
\begin{align}
C_{CDW} &= \left\langle n_{l} n_{0} \right\rangle \rightarrow n^2 
-\frac{K_\rho}{(\pi l)^2}  
+  B_{CDW} \frac{\cos(2k_F l)}{\left| l \right|^{K_{\rho}+1}} \nonumber \\
&\qquad\qquad\qquad\qquad\qquad+ {B'}_{CDW} 
\frac{\cos(4k_F l)}{\left| l \right|^{4 K_{\rho}}} \nonumber \\
C_{SDW} &= \left\langle S_{l}^z S_{0}^z \right\rangle 
\rightarrow -\frac{1}{(\pi l)^2} + 
B_{SDW} \frac{\cos(2k_F l)}{\left| l \right|^{K_{\rho}+1}} \ ,
\end{align}
while the pairing correlations exhibit an unmodulated asymptotic
behavior with an exponent $1+1/K_{\rho}$
\begin{align}
C_{SS} &= \left\langle c_{l,\uparrow}^\dag c_{l,\downarrow}^\dag 
c_{0,\downarrow} c_{0,\uparrow} \right\rangle \rightarrow 
\frac{B_{SS}}{\left| l \right|^{1/K_{\rho}+1}} \nonumber \\
C_{TS} &= \left\langle c_{l+1,\downarrow}^\dag 
c_{l,\downarrow}^\dag c_{0,\downarrow} c_{1,\downarrow} 
\right\rangle \rightarrow \frac{B_{TS}}{\left| l \right|^{1/K_{\rho}+1}} \ .
\end{align}
\end{subequations}
The amplitudes $A_{\xi},B_{\xi}$ depend also on the interactions and on the
filling. The shot noise obtained from this system is shown in
Fig.~\ref{fig:metal}. In contrast to the previous examples of
insulating phases, the momentum distribution [Fig.~\ref{fig:metal}(d)]
now has an infinite slope singularity at $q=k_{F}$, as is
characteristic of a TL liquid~\cite{Ogata90}. 
This pronounced feature can also be
recognized in the noise correlations around the Fermi points, even for
strong interaction strength.  Although particle-hole fluctuations (which are
always negative) are dominant for all cases of repulsive interactions,
we expect pairing correlations (with positive contributions) to be
visible as well, but with a weaker dependence on the system size. Our
numerical results confirm these expectations: In
$G_{\uparrow\uparrow}$, Fig.~\ref{fig:metal}(a) and $G_{\uparrow\downarrow}$,
Fig.~\ref{fig:metal}(b), the dips along
$k'=k\pm2k_{F}$ are well-pronounced, whereas the particle-particle
peaks close to the Fermi points are barely discernible. It is interesting
to note that the main features describing this metallic phase already
emerge from a simple perturbative calculation; see
appendix~\ref{sec:perturbation} and compare with
Fig~\ref{fig:metal-perturbation}. The behavior around opposing Fermi
points ($k=k_{F}$, $k'=-k_{F}$) is in perfect qualitative agreement
with recent bosonization results [Eq.~(8) of Ref.~\onlinecite{Mathey05}].  
The detection of 4$k_F$ charge
fluctuations, which are present in principle, is hindered by the fact
that they decay
as a power law with an exponent significantly larger than
two and presumably have a small amplitude $A'_{CDW}$.  

\subsection{Summary}
Let us now summarize the main signatures of
different phases that can be observed in the noise
correlations. 
As a general rule, particle-particle
fluctuations give a positive contribution and are observed along the
anti-diagonal $k'=-k$ of the shot noise, while particle-hole
correlations enter with a negative sign and lead to dips along $k'=k\pm q$. 
The strong signal along the diagonal $k'=k$ 
contains only information that 
is already encoded in the
momentum distribution. Tab.~\ref{tab:summary} summarizes the five
different phases encountered in the examples presented above,
indicating the sign of the singularities in the rescaled noise
correlations $LG$, together with the channels in which they can be
observed. 
\begin{table}
\centering
\caption{\label{tab:summary}
\emph{Diverging behavior observed in the rescaled noise correlations
  $LG$ for different phases. A plus (+) or minus (-) symbol indicates
  the sign of the singularity.}} 
\begin{tabular}{c|c|c}
Phase & $L G_{\uparrow\uparrow}$ & $L G_{\uparrow\downarrow}$\\ \hline
Charge-density wave (CDW)& - &0  \\
Spin-density wave (SDW) & - & -\\
Singlet superconductivity (SS) & 0 &+ \\
Triplet Superconductivity (TS) & +&+ \\
Bond-order wave (BOW) / dimerization &- & 0\\
\end{tabular}
\end{table}
If there is only one ordering tendency in the system, it can be
identified unambiguously. However, as soon as there are two or more
competing phases, one has to include additional information to 
distinguish between them.

\section{Experimental considerations\label{sec:experiments}}
Finally, 
we would like to put these numerical results in context with respect to
experiments. Initially, the atoms would be confined in an optical
lattice. By using PBC in our numerical calculations, we have neglected the
presence of a shallow confining potential as well as the 
open ends of the 1D system. 
We have tested the sensitivity of the
shot noise to the choice of boundary conditions in the
case of a combined CDW/SDW; see Sec.~\ref{sec:SDWCDW}. 
Qualitatively, we find exactly the same features for OPC as for 
PBC, namely, a pronounced dip along $k'=k\pm4 k_{F}$ in
$G_{\uparrow\uparrow}$, indicating the presence of a CDW, and a less
distinct valley along $k'=k\pm2 k_{F}$ in $G_{\uparrow\uparrow}$ and
$G_{\uparrow\downarrow}$, reflecting the SDW ordering (not
shown). After the trap is turned off, the
atom cloud expands
freely,
allowing the momentum distribution
$n_{\sigma}\left(k\right)$ as well as the noise correlations
$G_{\sigma\sigma'}\left(k,k'\right)$ to be measured. 
The relationship between the freely expanding 
atom cloud and the initial lattice states of the trapped
atoms is explained in detail in Ref.~\onlinecite{Altman04}. We are
thus confident that our results directly apply to experimental
realizations and measurements. The summary of features present in the
noise correlations (Tab.~\ref{tab:summary}) clearly illustrates the
necessity for state-selective measurements to distinguish
different phases, as the sign of the signal reveals only the character
of the fluctuation, i.e., particle-hole- or particle-particle-like.

\begin{figure*}
\includegraphics[width=0.8\textwidth,clip]{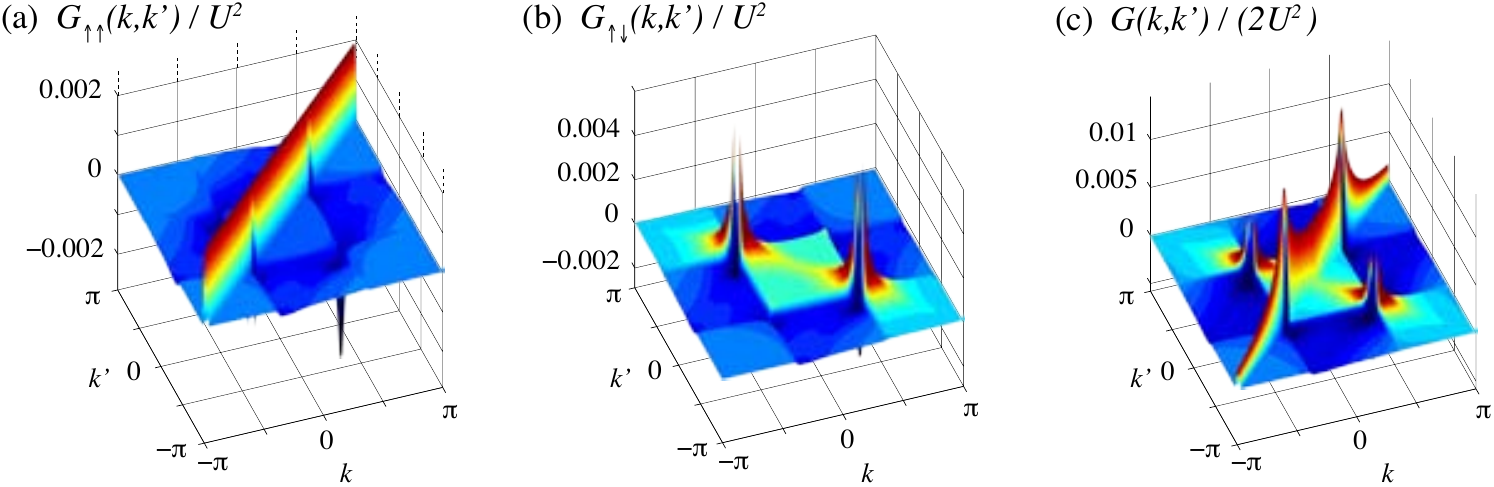}
\caption{\emph{(Color online). Noise correlations $G/U^2$ obtained in
    second-order perturbation theory for a metallic system at a filling
    of $x=3/8$. When compared to the DMRG results shown in
    Fig.~\ref{fig:metal}, the positive peaks in the vicinity of
    opposite Fermi points due to particle-particle correlations are
    more pronounced. The analytical expressions have been evaluated
    here for $L=56$ and $N=42$.
    }\label{fig:metal-perturbation}}
\end{figure*}

\section{Conclusion\label{sec:conclusion}}
We have analyzed the noise correlations (density-density
correlations in momentum space) in various phases of the extended
Hubbard model in one dimension. Our numerical density-matrix
renormalization group study, carried out for several characteristic
interactions and fillings, shows that different types of
(quasi-)long-range order leave 
different fingerprints in the shot noise. This allows 
different phases to be identified and distinguished with a universal
probe. 
The method is therefore
of interest for experiments with ultracold atoms, in which the
shot noise can be extracted from time-of-flight images. We have also
pointed out the importance of sum rules, i.e., the fact that the integral of
over the Brillouin zone is equal to zero, which leads to the interesting effect
that noise features due to correlations must be compensated by a (possibly
nonsingular) complementary structure.
In the future, it would be desirable to obtain similar ab-initio results for the noise
correlations of various fermionic systems in two dimensions, notably
the repulsive Hubbard model at and away from half filling.
Furthermore the effect of finite temperature on the noise profiles needs to 
be investigated.

\acknowledgments
We are grateful to O.~Sushkov and T.~Giamarchi for valuable
discussion. This work was supported by the Swiss National Science
Foundation. The ED calculations were performed on the Cray XT3 at
CSCS  (Manno, Switzerland). One of us (A.M.L.) acknowledges the warm
hospitality of the School of Physics at UNSW (Sydney, Australia)  during 
his stay in the final stage of the project.
\appendix

\section{Perturbative regime} \label{sec:perturbation}
For $t'=V=0$ and small on-site interactions $\left|U\right| \ll 1$,
one can calculate the noise correlations in perturbation theory. In
the momentum representation, the Hamiltonian~(\ref{eq:H}) reads
\begin{equation} \label{eq:h2}
H = \sum_{k,\sigma} \epsilon_{k} c_{k,\sigma}^\dag c_{k,\sigma} 
- \frac{U}{L}\sum_{\substack{k,k',q\\k \ne k'}} 
c_{k,\uparrow}^\dag c_{k',\downarrow}^\dag c_{q,\uparrow} 
c_{k+k'-q,\downarrow} \ ,
\end{equation}
with the dispersion
\begin{equation}
\epsilon_{k} = -2 \cos k + 2 \cos k_{F} 
\end{equation}
and Fermi momentum $k_{F}=x \pi$. To first order, the ground state
wave function is given by 
\begin{equation}
\left| \Psi \right\rangle = \left| F \right\rangle 
+ \frac{U}{L} \sum_{\substack{k,k',q\\k \ne k'}} 
\frac{c_{k,\uparrow}^\dag c_{k',\downarrow}^\dag c_{q,\uparrow} 
c_{k+k'-q,\downarrow}}{\epsilon_{k}+\epsilon_{k'}
-\epsilon_{q}-\epsilon_{k+k'-q}} \left| F \right\rangle \ , 
\end{equation}
where $\left|F\right\rangle=\prod_{\substack{k\leq k_{F}\\\sigma}}
c_{k,\sigma}^\dag \left|0\right\rangle$ is the Fermi sea.

\subsection{Momentum distribution}
At zero temperature, the momentum distribution of the noninteracting
system reads
\begin{equation*}
n^{(0)}_{\sigma}\left(k\right) = f\left(k\right) 
= \theta\left(\left|k\right|-k_{F}\right) \ ,
\end{equation*}
where $\theta\left(k\right)$ is the unit step function. 
For convenience, we also define $\bar f\left(k\right)=1-f\left(k\right)$.
The first corrections arise in second order perturbation theory and 
are therefore independent of the sign of $U$
\begin{equation}
n^{(2)}_{\sigma}\left(k\right) \approx
n^{(0)}_{\sigma}\left(k\right) 
+ U^2 \left[I_{1}\left(k\right)-I_{2}\left(k\right)\right] \ ,
\end{equation}
with
\begin{align*}
I_{1}\left(k\right)&=\frac{\bar f\left(k\right)}{L^2} 
\sum_{\substack{k',q\\q\ne k}}
\frac{\bar f\left(k'\right) f\left(q\right) 
f\left(k+k'-q\right)}{\left(\epsilon_{k}+\epsilon_{k'}
-\epsilon_{q}-\epsilon_{k+k'-q}\right)^2} \\
I_{2}\left(k\right)&= \frac{f\left(k\right)}{L^2} 
\sum_{\substack{k',q\\q\ne k}} 
\frac{\bar f\left(k'\right) \bar f\left(q\right) 
f\left(k'+q-k\right)}{\left(\epsilon_{k'}
+\epsilon_{q}-\epsilon_{k}-\epsilon_{k'+q-k}\right)^2} \ .
\end{align*}

\subsection{Noise correlations}
For \emph{non-interacting} fermions on a chain with PBC,
$G_{\sigma\sigma'}\left(k,k'\right) \equiv 0$ because $n_{k,\sigma}$
commutes with the tight-binding Hamiltonian. Analogous to the momentum
distribution, the lowest order corrections to the noise correlations
are quadratic in $U$ and thus also independent of the sign
\begin{widetext}
\begin{equation}
G^{(2)}_{\sigma\sigma'}\left(k,k'\right) \approx U^2
\begin{cases}
\delta_{\sigma,\sigma'}\delta_{k,k'} I_{1}\left(k\right) 
+ \left(1-\delta_{\sigma,\sigma'}\right) I_{4}\left(k,k'\right) 
- I_{1}\left(k\right) I_{1}\left(k'\right) & \left(k>k_{F} 
\wedge k'>k_{F}\right) \\
- I_{3}\left(k,k'\right) + I_{1}\left(k\right) 
I_{2}\left(k'\right) & \left(k>k_{F} \wedge k'\leq k_{F}\right) \\
\delta_{\sigma,\sigma'} \delta_{k,k'} I_{2}\left(k\right) 
+ \left(1-\delta_{\sigma,\sigma'}\right) I_{5}\left(k,k'\right)
- I_{2}\left(k\right) I_{2}\left(k'\right) & \left(k\leq k_{F} 
\wedge k'\leq k_{F}\right) 
\end{cases}
\end{equation}
\end{widetext}
with the two-point correlations $I_{j}$ given by
\begin{align*}
I_{3}\left(k,k'\right)&=\frac{\bar f\left(k\right) f\left(k'\right)}{L^2} \sum_{q} 
\frac{\bar f\left(q\right) f\left(k+q-k'\right) }{\left(\epsilon_{k}+\epsilon_{q}-\epsilon_{k'}-\epsilon_{k+q-k'}\right)^2} \ , \\
I_{4}\left(k,k'\right)&= \frac{\bar f\left(k\right)\bar f\left(k'\right)}{L^2} \sum_{q} 
\frac{f\left(q\right)  f\left(k+k'-q\right) }{\left(\epsilon_{k}+\epsilon_{k'}-\epsilon_{q}-\epsilon_{k+k'-q}\right)^2} \ , \\
I_{5}\left(k,k'\right)&= \frac{f\left(k\right) f\left(k'\right)}{L^2} \sum_{q} 
\frac{\bar f\left(q\right)\bar f\left(k+k'-q\right)}{\left(\epsilon_{k}+\epsilon_{k+k'-q}-\epsilon_{k}-\epsilon_{k'}\right)^2} \ .
\end{align*}

The normalized noise correlations $G/U^2$ obtained in this
perturbation calculation are shown in
Figs.~\ref{fig:metal-perturbation} for a metallic system at filling
$x=3/8$. Compared to the DMRG results shown in Fig.~\ref{fig:metal},
the positive peaks in the vicinity of opposite Fermi points due to
particle-particle correlations are more pronounced, but, otherwise, the
perturbative approach leads to a surprisingly accurate description of
the shot noise.

\end{document}